\documentclass[%
 reprint,
 amsmath,amssymb,
 aps,
pra,
nofootinbib,
]{revtex4-1}

\usepackage{graphicx}
\usepackage{dcolumn}
\usepackage{bm}
\usepackage{physics}
\usepackage[normalem]{ulem}

\usepackage{xcolor}
  \usepackage{amsmath,amssymb}
\usepackage{dblfloatfix}
\usepackage{booktabs}

\DeclareMathOperator{\aop}{\hat{a}}
\DeclareMathOperator{\cop}{\hat{a}^\dagger}
\DeclareMathOperator{\delop}{\hat{\delta}}

\DeclareMathOperator{\ea}{\expval{\aop}}

\DeclareMathOperator{\ph}{\theta}
\DeclareMathOperator{\phop}{\hat{\ph}}

\DeclareMathOperator{\nop}{\hat{n}}
\DeclareMathOperator{\en}{\ev{\nop}}
\DeclareMathOperator{\enn}{\expval*{\nop\nop}}
\DeclareMathOperator{\dn}{\hat{\delta}_n}
\DeclareMathOperator{\dph}{\hat{\delta}_{\ph}}
\DeclareMathOperator{\edndn}{\ev*{\dn\dn}}
\DeclareMathOperator{\eph}{\ev*{\phop}}
\DeclareMathOperator{\ephph}{\ev*{\phop\phop}}
\DeclareMathOperator{\edphdph}{\ev*{\dph\dph}}
\DeclareMathOperator{\enph}{\ev*{\nop\phop}}

\DeclareMathOperator{\edndphsym}{\ev*{\dn\dph}_{\text{sym}}}
\DeclareMathOperator{\Var}{Var}
\DeclareMathOperator{\mean}{Mean}
\DeclareMathOperator{\intravar}{\Var_1}
\DeclareMathOperator{\intervar}{\Var_2}

\DeclareMathOperator{\Oop}{\hat{O}}

\DeclareMathOperator{\ham}{\hat{H}}

\newcommand{\wv}[1]{#1}

\newcommand{\mw}[1]{#1}
\newcommand{\mwsc}[1]{}
\newcommand{\mwsout}[1]{}
\renewcommand{\ol}[1]{\overline{#1}}

\begin{document}

\preprint{APS/123-QED}

\title{The temporal coherence of a photon condensate: A quantum trajectory description}

\author{Wouter Verstraelen}
\email{wouter.verstraelen@uantwerpen.be}
\author{Michiel Wouters}%
 \email{michiel.wouters@uantwerpen.be}
\affiliation{%
Theory of Quantum \& Complex Systems, University of Antwerp, B-2610 Wilrijk, Belgium
}%

\date{\today}

\begin{abstract}
In order to study the temporal coherence of a single-mode dye-cavity photon condensate, a  model is developed for the dynamics which treats the condensate mode on a quantum-mechanical level. The effects of driving-dissipation and Kerr interactions on the number fluctuations are studied analytically and numerically, including the finding of a long-$\tau$ antibunching effect.  Depending on the interaction strength, we quantitatively observe an exponential Schawlow-Townes-like decay or Gaussian Henry-like decay of phase correlations. The adequacy of a heuristic phasor model originating from laser physics in describing number and phase dynamics is validated within the experimentally relevant parameter regime. The ratio of the first and second order coherence times is shown to be inversely proportional to the number fluctuations, with a prefactor that varies smoothly throughout the crossover between canonical and grandcanonical statistics.
\end{abstract}

\pacs{Valid PACS appear here}
\maketitle


\section{Introduction}

The last few decades, a vast amount of research has been done to the related phenomena of Bose-Einstein Condensates (BECs) and quantum fluids~\cite{pitaevski_stringari_2016,bramati_modugno_2015}, for reasons of fundamental interest as well as their exploitation for quantum simulation. 

The traditional platform for the realization of quantum fluids are ultracold atoms, but another fascinating platform is a BEC of photons. Because their effective mass is many orders of magnitude lower, condensation can take place at room temperature. In order to achieve thermalization, a successful approach has been to use effective photon-photon interactions in a nonlinear material, resulting in the condensation of polaritons: coherent superpositions of photons and excitons~\cite{qfloflight}, of which the temporal coherence has been studied in \cite{Whittakerpolaritons1,*Whittakerpolaritons2,HaugPolaritonCoherence}.

An alternative approach, where there is no need for the photons to be hybridized, consists of letting the photons interact with their environment~\cite{originalnatureKlaers2010,originalnatphysKlaers2010,becreviewRajan2016,photonBECsmalltolargedoi:10.1080/09500340.2017.1404655,KirtonKeelingThermalizationPhysRevA.91.033826,vanOostendichtheidsv,nonmarkovsimulation}. This is realized experimentally by frequent absorptions and emissions of the photons by dye-molecules (the gain medium), that are themselves thermalized by collisions with solvent molecules. It has been shown ~\cite{KirtonKeelingCrossoverPhysRevLett.111.100404,SchmittCrossoverPhysRevA.92.} that there is a smooth crossover between laser physics and photon BEC. The distinction is that in the latter, almost full thermalization is reached with minor leakage, whereas in a laser, which is driven further out-of-equilibrium, gain and losses dominate the dynamics. A consequence is that lasers generally need population inversion for the gain medium, while a photon BEC does not.

An interesting, presently realized, feature of photon BECs in a dye microcavity system, as opposed to other BECs, is that the number of photons does not remain fixed through time, even in the absence of losses. If the gain medium is sufficiently large such that saturation effects are small, the condensate will exhibit grand-canonical statistics \cite{KlaerststatphysofBECPhysRevLett.108.160403,JulianFlickeringPhysRevLett.112.030401,JacquesGrandCanonicalPhysRevE.94.042124,proukakis_snoke_littlewood_2017}. This is remarkable, because the non-interacting Bose gas is one of the few systems where ensemble equivalence is not satisfied \cite{KersonHuang_2014}.

Along with these statistical fluctuations of the particle number, one observes fluctuations of the phase \cite{Schmittphasecoherence,thesisJulian,*thesisJulianEng,Spatiotemporal_Multimode}, as first predicted by \cite{deLeewPhasediffusion}. These can be understood from a heuristic phasor model (HPM), originally proposed in laser physics \cite{henry,scully_zubairy_1997}, because the phase fluctuations at sufficient particle number correspond to the standard Schawlow-Townes broadening \cite{scully_zubairy_1997}. The dynamics seems richer though, because in the grand-canonical regime the particle number can become zero so that the phase of a subsequent new photon is entirely random, known as a `phase jump'. Although predictions from the HPM seem to match experiment so far, the theoretical understanding is still limited.

In particular, the phase jump picture suggests that a dramatic increase of temporal coherence occurs when the second order coherence falls below roughly $g^{(2)}(0)=1.5$, where the probability of having zero photons in the condensate starts to become exponentially small~\cite{Schmittphasecoherence}. We will show here that throughout the crossover between canonical and grandcanonical statistics, the behaviour of the phase coherence time behaves much smoother than expected by the phase jump analysis.

We also study the influence of a weak Kerr nonlinearity \mwsout{effect} on the \mw{coherence}\mwsout{dynamics}. For ultracold atoms, the temporal coherence properties in the presence of interactions have been studied in \cite{BECcoherence} for bosons and \cite{HadrienFermiFR} for fermions. Nonlinearities are not necessary for condensation to occur and are thus often disregarded \cite{thesisJulian,*thesisJulianEng}. Nevertheless, they naturally emerge in experimental setups, where a not entirely decoherent dye \mw{induces} \mwsout{facilitates} a natural Kerr effect~\cite{InterplayPelster}. Furthermore, it is possible to engineer these interactions on purpose, which makes the photons behave more like polaritons and might open the possibility for effects such as superfluidity \cite{superfluidphotons}.  It has also been proposed to realize similar behavior with $\chi^{(2)}$-nonlinear materials \cite{chi2}. Although the precise \mw{intrinsic} \mwsout{natural} value \mw{of the Kerr nonlinearity in the commonly used dye molecules} is still subject of debate  (a recent discussion is given in \cite{InterplayPelster}), we show that already small values of the Kerr nonlinearity can \mw{subsantially} \mwsout{strongly}  affect the coherence properties. Because the interaction strength can vary many orders of magnitude, we here focus on the qualitative influence on interaction scales where the number and phase dynamics are altered significantly, while the associated energy scale per particle remains small compared to other energy scales.
Apart from the instantaneous Kerr effect, delayed interaction may also arise in photonic condensates because of thermo-optical effects \cite{ThermoOpticalNJP}. The latter are out of the scope here because they are expected to contribute significantly only on timescales much longer than our simulations, although in principle our model could be extended to include them.

\mwsc{Moved:[}
Throughout this work, we will assume for simplicity the presence of only a single photon mode. In many experimental setups, this is valid because of the lower population of excited modes, the fact that these do not interact and their \mw{small} \mwsout{low} overlap with the condensate \cite{applphysBkerr}. Analytical predictions considering only a single mode have also been successful in describing in phenomena as the decay of second-order coherence \cite{thesisJulian,*thesisJulianEng}. As confinement is improving with novel experiments, restriction to single mode becomes more realistic in practice. 
Nevertheless, \mw{beyond} this single-mode description, photonic condensates in some parameter regimes can also exhibit effects arising from multiple modes that \mw{lead to} interesting physics: if the occupation of other modes is large and the finite size of the reservoir is important, mode-competition can become important \cite{nonmarkovsimulation}. Large occupation of other modes in a substantially interacting system, will additionally result in Beliaev-Landau scattering which may give corrections to the single-mode results \cite{pitaevski_stringari_2016}.\mwsc{]}

In the next section, Sec. \ref{sec:numbers}, we describe the photon condensate system in more detail, and study the number statistics. We revisit standard results with inclusion of a Kerr \mw{nonlinearity} and present thermodynamic and dynamical \mw{estimates} \mwsout{arguments} for the number fluctuations, which are verified by solving stochastic rate equations. We also show that weak interactions can alter significantly the density distribution and intensity correlations; and further that finite losses can cause long-$\tau$ antibunching.

In Sec \ref{sec:HPM}, we \mw{tackle} \mwsout{revisit} the semiclassical HPM in presence of Kerr interactions,  and point out some conceptual issues stemming from its the heuristic nature.

To verify the validity of the HPM, we provide in Sec. \ref{sec:TD} a model where the condensate mode is treated on a fully quantum-mechanical level, while the dye molecules are described by classical rate equations. As we aim to describe single-shot experimental realizations, this naturally leads to the quantum trajectory (wave-function Monte Carlo) formalism yielding a stochastic Schr\"{o}dinger equation that describes an open system under continuous measurement \cite{breuer,carmichaelbook}.  Because photon numbers in the cavity can easily reach order $10^4$, it is numerically very demanding to solve this Schr\"{o}dinger equation exactly on a truncated Fock space. Therefore, at high densities, we make the variational ansatz that the state is Gaussian in density and phase, which has shown to be good in describing dephasing properties \cite{gausmethode}. In addition to reducing computational complexity, the variational equations also yield theoretical insight in the density-phase dynamics. Quantum trajectories have previously also been used for the study of the related polariton condensates \cite{polaritontrajectories}.

Using this quantum trajectory model, we show that, despite its heuristic character, the HPM is valid for experimentally relevant quantities as their predictions from both methods match.
\mw{Finally}, In Sec. \ref{sec:1coherence} we \mwsout{finally} study the properties of first order coherence in photon condensates, including the influence of Kerr interactions and reservoir size, and relate \mw{the first and second order coherences to each other} \mwsout{them to the second order coherence}. Both Schawlow-Townes and Henry phase diffusion mechanisms are observed, and we show that the presence of phase jumps only \mwsout{has} a limited quantitative \mw{effect on} the phase decay. 
\mw{We formulate our conclusions} in Sec. \ref{sec:conclusions}.

\section{Number statistics}
\label{sec:numbers}
\subsection{\mw{Model for a driven-dissipative, interacting photon condensate}}
We assume a reservoir containing $M_{\text{tot}}$ two-level dye molecules, of which $M_\uparrow (M_\downarrow)$ are in the excited state (ground state). Because of the rapid decoherence due to collisions with solvent molecules \cite{KirtonKeelingCrossoverPhysRevLett.111.100404}, these numbers can be treated as classical integers. We consider this reservoir to be coupled with a single, quantum mechanical, photon mode with number operator $\nop$. 
Because we restrict ourselves in this section to number statistics, it is sufficient to consider only number eigenstates (Fock states) with $n$ photons. The corresponding rate equation for the photons is then given by
\begin{equation}
dn=-dN+dM.\label{eq:rateeqsimple}
\end{equation}

Here $dN$ and $dM$  are Poisson processes describing molecular absorption at rate $B_{12}M_{\downarrow}n$ and molecular emission at rate ${B_{21}M_{\uparrow}(n+1)}$, respectively, where $B_{12}$ and $B_{21}$ are the modified Einstein coefficients for absorption and emission \cite{KlaerststatphysofBECPhysRevLett.108.160403}. For convenience, we define $\gamma:=B_{12}M_\downarrow$ and $R:=B_{21}M_{\uparrow}$.  In \mw{the absence of losses,} \mwsout{first approximation} the total number of excitations $X=M_{\uparrow}+n$ is \mwsout{considered} conserved such that $dM_{\uparrow}=-dn$ and $dM_{\downarrow}=dn$. For this system, many previous analytical results regarding the number distribution and correlation functions have been summarized in \cite{thesisJulian,*thesisJulianEng}. 
In practice, the quality factors of cavities are restricted so that an additional loss at rate $\kappa n(t)$ \cite{originalnatureKlaers2010} takes place, which is compensated by an additional pumping of molecules. In a typical experimental setup,  $\gamma\gg\kappa$, so that we expect losses not to affect the number statistics significantly. As we are concerned with ensemble statistics only, we have the freedom to choose the photon-counting unraveling for external losses \cite{quantumnoise}, such that they are also modelled by a Poisson process $dK$. This results in a rate equation

\begin{equation}
dn=-dN+dM-dK\label{eq:rateeqwithlosses}.
\end{equation}
In order to keep the long-time expectation value of $X$ constant, the molecule reservoir is pumped through a process $dP$ with constant rate $\kappa\ol{n}$, where
$\ol{n}$ is the average particle number such that 
\begin{equation}\label{eq:rateeqmols}
dM_{\uparrow}=dN-dM+dP 
\end{equation}
 and $M_{\downarrow}=M_{tot}-M_{\uparrow}$.
From equation \eqref{eq:rateeqwithlosses},  the average evolution is  \cite{thesisJulian,*thesisJulianEng}
\begin{equation}
\pdv{n}{t}=B_{21}(X-n)(1+n)-B_{12}n(M-X+n)-\kappa n,
\end{equation}
from which one finds that in the steady state
\begin{equation}\label{eq:Xfromn}
X(\ol{n})=\frac{\ol{B_{12}n(M+n)+B_{21}n(n+1)+\kappa n}}{\ol{B_{12}n+B_{21}(n+1)}}.
\end{equation}
In general, the Einstein-coefficients $B_{12},B_{21}$ of the absorption and emission processes are related by the Kennard-Stepanov law (see the supplementary material of \cite{KlaerststatphysofBECPhysRevLett.108.160403}):
\begin{equation}\label{eq:ksgeneral}
\frac{B_{21}(\omega)}{B_{12}(\omega)}=\frac{w_\downarrow}{w_\uparrow}e^{-\beta(\omega-\omega_0)}.
\end{equation}Here, $\omega$ is the frequency of the photon, $\omega_0$ the frequency of the atomic transition between ground and excited state of the dye, $\beta=\frac{1}{T}$ the inverse temperature (we set $k_B=\hbar=1$ throughout this work) and \mw{$w_{\uparrow,\downarrow}$} \mwsout{$w_\downarrow$, $w_\downarrow$} are weight coefficients taking into account the internal molecular density of (rovibrational) states in the electronic ground and excited level.
For \mwsout{a single mode of} noninteracting photons with frequency $\omega_c$, \mw{the  frequency dependence in \eqref{eq:ksgeneral}  depends only on the detuning  $\omega_c-\omega_0=\Delta$.}

A Kerr effect adds by definition an interaction energy $E_\text{int }(n)=\frac{U}{2}n^2$ to the photons. The $n$th photon now carries a frequency $\omega=\omega_c+[E_\text{int}(n)-E_\text{int}(n-1)]\approx\omega_c+Un$.  As long as the interaction strength and particle number are not excessively large, only the relative difference between $B_{21}(\omega)$ and $B_{12}(\omega)$ is important, so that we can treat $B_{12}$ as constant. With this, we rewrite \eqref{eq:ksgeneral} as
\begin{equation}\label{eq:ksinteracting}
\frac{B_{21}(n)}{B_{12}}=\frac{w_\downarrow}{w_\uparrow}e^{-\beta(\Delta+Un)}.
\end{equation}

\subsection{Number statistics}

As in  Ref.~\cite{KlaerststatphysofBECPhysRevLett.108.160403} the steady state number distribution can be found by assuming detailed balance 
\begin{equation}
\frac{\mathcal{P}_{n+1}}{\mathcal{P}_n}=\frac{X-n}{M-X+n+1}\frac{B_{21}(n)}{B_{12}}
\end{equation}
such that
\begin{equation}\label{eq:probabgen}
    \frac{\mathcal{P}_n}{\mathcal{P}_0}=\frac{\binom{X}{n}}{\binom{M_{\text{tot}}-X+n}{n}}e^{-\beta\left[\Delta n+(U/2)n^2\right]},
\end{equation} where the notation between brackets refers to binomial coefficients. 
In the limit of an infinite reservoir, \eqref{eq:probabgen} reduces to a Bose-Einstein distribution
\begin{equation}\label{eq:probabBEC}
\mathcal{P}_n\propto\left(\frac{\ol{M_{\uparrow}}}{\ol{M_{\downarrow}}}\right)^ne^{-\beta E[n]}=\exp\left(-\frac{(\Delta-\mu)n+(U/2)n^2}{T}\right),
\end{equation}
where the chemical potential is $\mu=T\log{\left(\ol{M_{\uparrow}}/\ol{M_{\downarrow}}\right)}$.

\begin{figure}
    \includegraphics[width=\columnwidth]{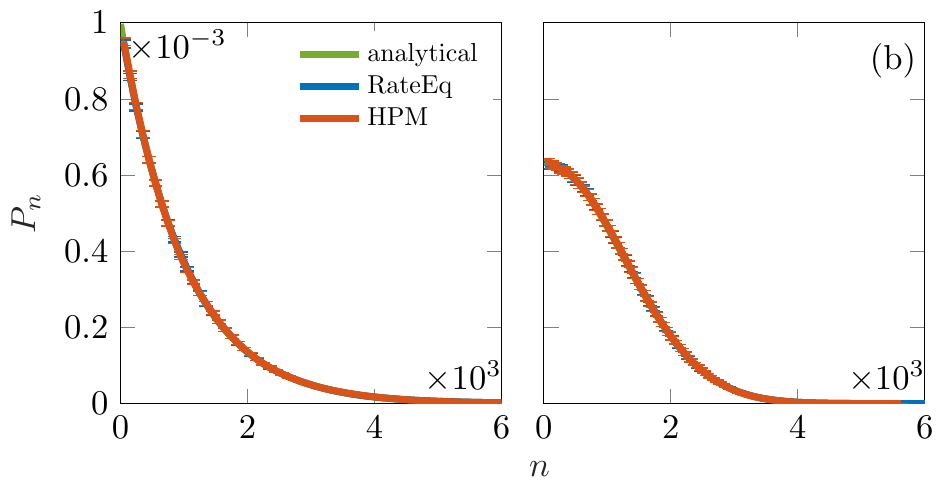}
    \caption{Particle number distribution \eqref{eq:probabgen} without (a) and with (b) interactions, for the parameters given in Tab. \ref{tab:parameters}. Green lines: analytical results \eqref{eq:probabgen}, covered by the numerical simulation of the rate-equations \eqref{eq:rateeqwithlosses}, \eqref{eq:rateeqmols} (blue) and HPM (red) results. In both cases there is a very good agreement between the analytical prediction and numerical results obtained by rate equations. These are also matched by the predictions from the HPM. Stochastic results are obtained from $10^3$ independent samples, each evolving a time $10^5 B_{12}^{-1}M_\text{tot}^{-1}$.}
    \label{fig:densitydistributions}
\end{figure}

In Fig. \ref{fig:densitydistributions}, we see that the predictions in number distributions \eqref{eq:probabgen}, closely match numerical results from rate equations \eqref{eq:rateeqwithlosses}, for a set of parameters corresponding to recent experiments \cite{parameters}, as summarized in Tab. \ref{tab:parameters}.
As for the interaction strength, we have taken the value $U=10^{-5}\hbar M_{\text{tot}}B_{12}$, corresponding to a dimensionless interaction parameter $\tilde{g}=\frac{U}{\Omega\hbar}\approx 9.95\times 10^{-5}$ for trapping frequency $\Omega\approx 8\pi\times 10^{10}~Hz$, which is larger than the most common estimates of the natural Kerr effect \cite{applphysBkerr,InterplayPelster}. Importantly, by defining the effective reservoir size 
\begin{equation}\label{eq:meff}
M_{\text{eff}}=\frac{M_{\text{tot}}}{2}\left[1+\cosh\left(\beta(\Delta+U\ol{n})\right)\right]^{-1},
\end{equation}
one can distinguish \mw{in the noninteracting case} a canonical regime with Poissonian number statistics ($\ol{n}^2\gg M_{\text{eff}}$), a grandcanonical regime with Bose-Einstein statistics ($\ol{n}^2\ll M_{\text{eff}}$) and a transition region ($\ol{n}^2\approx M_{\text{eff}}$). 

\begin{table}
\centering
\begin{ruledtabular}
\begin{tabular}{lr}
Parameter    & value  \\ \midrule\hline
$B_{12}$      & 2.5~kHz        \\
$M_{tot}$       & $10^9$            \\
$\Delta$        & $-2.4 kT/\hbar$     \\
$T$		&  $300~K$ \\
$U$ & $10^{-5}~B_{12}M_{\text{tot}}=6.4\times 10^{-7}~k_B T/\hbar$ \\
$w_{\uparrow}/w_{\downarrow}$  & 1    \\
$\kappa$  & $2.2~GHz = 8.3\times 10^{-4}~B_{12}M_{\text{tot}}$ \\ 
\end{tabular}
\end{ruledtabular}
\caption{Experimental parameters corresponding to Rhodamin 6G at 560~nm. \cite{parameters}. Further, we have taken for the simulations $\ol{n}=1000$, which somewhat smaller than in typical experimental setups to reduce the relevant timescales. We have chosen the finite value of interaction strength such that the phase and number statistics are significantly altered ($\sigma=\frac{U\ol{n}^2}{T}$ of order one), while the average interaction energy per particle $U\ol{n}$ remains small compared to other energy scales. The notion of `without interactions' refers to $U=0$, whereas `without losses' means $\kappa=0$}.
\label{tab:parameters}
\end{table}

\begin{figure}
\includegraphics[width=\columnwidth]{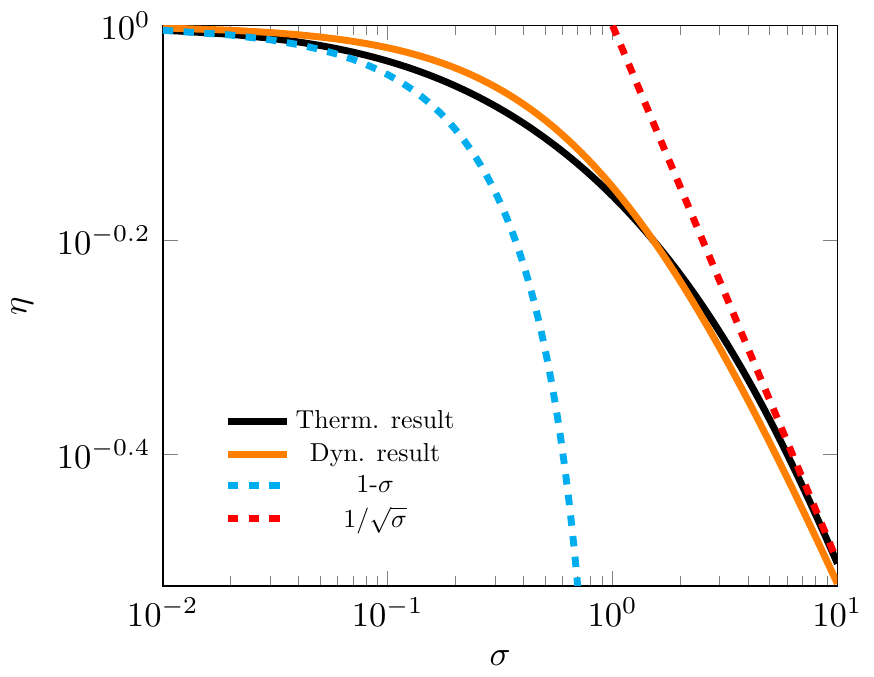}
\caption{Amount of fluctuations $\eta=\frac{\sqrt{\ol{n^2}-\ol{n}^2}}{\ol{n}}$ in a grandcanonical condensate as function of the interaction parameter $\sigma$ as thermodynamically derived (black), with asymptotics $1-\sigma$ (cyan, dashed) and $\sigma^{-1/2}$ (red, dashed). Orange: the dynamical approximate result \eqref{eq:alggtwo0}.
}
\label{fig:relflucts}
\end{figure}

For our parameters in Tab. \ref{tab:parameters}, we have an effective reservoir size $M_{\text{eff}}=7.6\times10^7\gg \ol{n}^2=10^6$ such that the system is rather on the grandcanonical side of this crossover. Definition \eqref{eq:meff} only weakly depends on $U$ to the extent that it changes the average energy per photon. However, also for fixed $M_\text{eff}$, the number distribution is significantly altered by interactions: the number fluctuations are reduced, as predicted by \cite{JacquesGrandCanonicalPhysRevE.94.042124}. This can be understood thermodynamically:  by approximating $n$ to be continuous and integrating \eqref{eq:probabBEC} over positive $n$, we obtain the grand-canonical partition function $\mathcal{Z}$. From the associated free energy $F=-T\log\mathcal{Z}$, the equation of state $\ol{n}(\mu)$ can be obtained, as wel as the amount of number fluctuations $\Var{[n]}(\mu)$. Eliminating $\mu$, the ratio $\eta=\frac{\sqrt{\ol{n^2}-\ol{n}^2}}{\ol{n}}$ can be expressed as a function of the interaction parameter $\sigma=\frac{U\ol{n}^2}{T}$, as shown on Fig. \ref{fig:relflucts}. Asymptotically, The relative amount of fluctuations $\eta$ decreases as $1-\sigma$ for small $\sigma$ and \mw{as} $\sigma^{-1/2}$ for large interactions $\sigma$. For our parameters, we find $\sigma=0.64$, corresponding to $\eta=0.75$, whereas our noninteracting condensate has $\eta=0.99$.

For a finite reservoir, $\mu$ becomes dependent on $n$, so that the above thermodynamic analysis is no longer valid. $\eta$ can also be computed self-consistently from the number distribution \eqref{eq:probabgen}, or analytically from the dynamical argument below.

\subsection{Second-order coherence}

\begin{figure}
\includegraphics[width=\columnwidth]{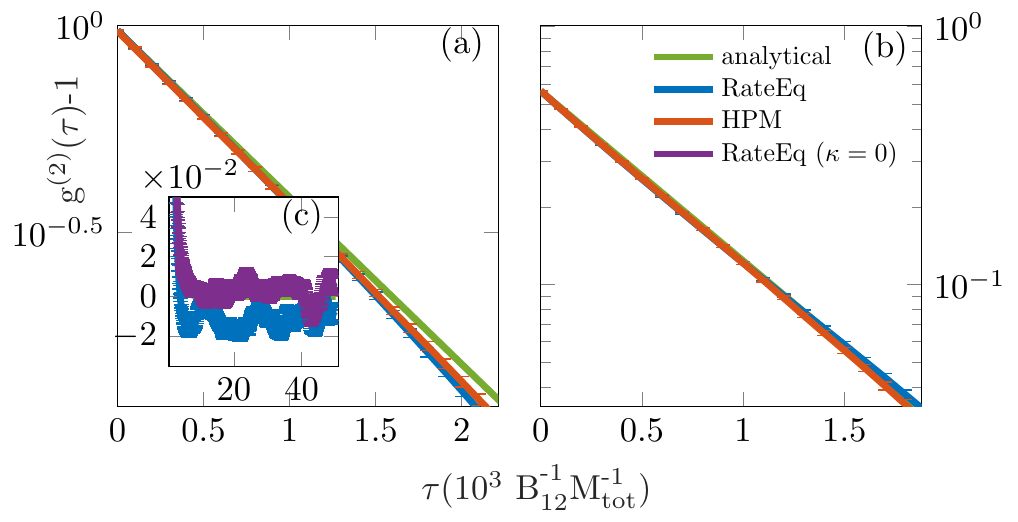}
    \caption{Second order correlation functions for the grandcanonical parameters of Tab. \ref{tab:parameters}, with the exponential decay \eqref{eq:interactingtc2}, without (a) and with (b) interactions. Generally there is a good agreement, though at later times deviations originating from the driven-dissipative character are visible. These results are again matched by the HPM (colors as in Fig. \ref{fig:densitydistributions}. Inset (c): $g^{(2)}(\tau_c^{(2)}<\tau<\tau_X)\wv{-1}$  becomes negative, while this is not the case without losses ($\kappa=0$, purple).}
    \label{fig:gtwos}
\end{figure}

The second order coherence time $\tau_c^{(2)}$, the decay time of $g^{(2)}(\tau)=\frac{\ol{n(t)n(t+\tau)}}{\ol{n}^2}$,  can be obtained by an extension of the approach in \cite{thesisJulian,*thesisJulianEng}. In linear approximation, $B_{21}=\ol{B_{21}}(1-\beta U\delta n(t))$, such that the average fluctuation evolves as 
\begin{align}
\pdv{t}&\delta n(t)=\\&-\left[\frac{\ol{B_{21}}X}{\ol{n}}+(B_{12}+\ol{B_{21}})\ol{n}+\beta U\ol{B_{21}}M_{\uparrow}(1+\ol{n})\right]\delta n(t)\nonumber\\&+\mathcal{O}(\delta n(t)^2) ,\nonumber
\end{align}
where the driven-dissipative nature has been disregarded.
Accordingly, the number correlations decay at rate
\begin{align}\label{eq:interactingtc2}
\Gamma_2&=\left[\frac{1+\sigma}{\ol{n}(1+e^{\beta(\Delta+U\ol{n})})}+\left(1+e^{-\beta(\Delta+U\ol{n})}\right)\frac{\ol{n}}{M_{\text{tot}}}\right]B_{12}M_{\text{tot}}\\
&=\left[1+\sigma+\frac{\ol{n}^2}{M_\text{eff}}\right]\frac{B_{12}\ol{M_\downarrow}}{\ol{n}},
\end{align}
As long as $U\ol{n}\ll\Delta$, the change in $\Gamma_2=(\tau_c^{(2)})^{-1}$ from a noninteracting condensate is directly proportional to the interaction parameter $\sigma$.  Note that the same expression \eqref{eq:interactingtc2} is also obtained if $B_{21}$ is treated as constant while considering frequency dependence in $B_{12}$.
We verify the validity of decay rate \eqref{eq:interactingtc2} by comparison to numerical simulations by rate equations in Fig. \ref{fig:gtwos}. 

For generic parameter values, $g^{(2)}(0)=1+\eta^2$ can also be estimated by the stationary solutions of the expectation value of
 \begin{align}
 d\left[\delta n(t)^2\right]&=2\delta n(t) d\left[\delta n(t)\right]+d\left[\delta n(t)\right]^2\\
 &=-2\Gamma_2\delta n(t)^2 dt+R(\ol{n}+\delta n+1)dt+\gamma (\ol{n}+\delta n),
 \end{align}
 from which
 \begin{equation}\label{eq:alggtwo0}
 g^{(2)}(0)-1=\eta^2=\frac{\ol{\delta n^2}}{\ol{n}^2}=\frac{1}{1+\sigma+\frac{\ol{n}^2}{M_\text{eff}}}.
 \end{equation}
 For an infinite reservoir ($M_\text{eff}/\ol{n}^2\rightarrow\infty$), we can compare this dynamical result directly with the thermodynamic result above. As we see \wv{in} Fig. \ref{fig:relflucts}, the relative difference in predictions for $\eta$ remains smaller than 5\% for all values of $\sigma$, and \eqref{eq:alggtwo0} becomes exact in the limit of large $\sigma$. We can attribute the deviations at smaller $\sigma$ to higher order contributions in $d\delta n$. 

Such an estimate of $\eta$ is also useful to initiate the number of excited molecules in the numerical simulations. As can be seen from Eq. \eqref{eq:Xfromn}, both the average number of photons and its fluctuations determine the number of excited molecules:
\begin{equation}
M_{\uparrow}=X(\ol{n})-\ol{n}=\frac{(B_{12}M+\kappa)\ol{n}+(B_{12}+B_{21})\ol{n}^2\eta^2}{B_{21}(\ol{n}+1-\beta U\ol{n}^2\eta^2)+B_{12}\ol{n}}.
\end{equation}

In previous analytic discussions, we have disregarded the driven-dissipative nature of the system. \mw{Intuitively, one might expect that losses will reduce the second order coherence time.}
\mwsout{It might be expected that the pumping at constant rate, where losses fluctuate, leads to a damping to the fluctuations,} \mw{Such} an effect is not observed in the number distribution, as seen in Fig. \ref{fig:densitydistributions}. 
\mwsc{In the meantime, I have done the analytical calculation in the presence of losses, maybe we can include this.}

However, as we see in Fig.\ref{fig:gtwos}, $g^{(2)}(\tau)$ is altered on timescales long compared to the photon correlation time $\tau_c^{(2)}$, \mw{where} \mwsout{as} a long-$\tau$ antibunching effect appears. 
This can be explained in the following way: if at $\tau=0$ the stochastic particle number is $n=\ol{n}+\delta n$, the corresponding fluctuation modifies the excitation number by $\Delta X=-\kappa\,\delta n\tau_c^{(2)}$. Because $\ol{n}=\ol{n}(X)$, the particle number expectation value is altered to  $\ol{n}'=\ol{n}+\pdv{\ol{n}}{X}\Delta X$  where from \eqref{eq:Xfromn}, $\pdv{X}{\ol{n}}\approx\frac{B_{12}B_{21}M_{tot}}{\ol{n}^2(B_{12}+B_{21})^2} $, 
so that $g^{(2)}(\tau)$ changes in the order of
\begin{equation}\label{eq:gtwochange}
\frac{\ol{n n'}}{\ol{n}^2}-\frac{\ol{n n}}{\ol{n}^2} \approx\frac{\ol{n}^2(B_{12}+B_{21})^2}{B_{12}B_{21}M_{tot}}\kappa\tau_c^{(2)},
\end{equation}
where we used that in the grandcanonical regime, $ \delta n\sim \ol{n}$.
Expression \eqref{eq:gtwochange} predicts a decrease in $g^{(2)}(\tau)$ of about $0.012$. We verify this numerically by averaging $g^{(2)}(\tau)$ over a time-interval $10^{4}B_{12}^{-1}M_\text{tot}^{-1}-5\times10^{4}B_{12}^{-1}M_\text{tot}^{-1}$, which is sufficiently larger than $\tau_c^{(2)}$ but smaller than the timescale of fluctuations in total excitation number $\tau_X$ and obtain  a value $g^{(2)}(\tau_c^{(2)}<\tau<\tau_X)-1=-0.012$ for the experimental $\kappa$ and $g^{(2)}(\tau_c^{(2)}<\tau<\tau_X)-1=0.003$ without losses, in agreement with our results. Of course, one always has $g^{(2)}(\infty)=1$ as initial and final state become entirely independent, but  the relaxation from deviation \eqref{eq:gtwochange} only takes place over the timescale of the dynamics of $X$, namely $\tau_X=(\kappa\pdv{\ol{n}}{X})^{-1}$, which is of order $10^5~B_{12}^{-1}M_{tot}^{-1}$ in our simulations. Because of the different values for $\tau_c^{(2)}$ and $\eta$, this effect is weaker in our \mw{simulations of the} interacting \mw{condensate}\mwsout{case}.

\section{The Heuristic Phasor Model for phase evolution}
\label{sec:HPM}

\subsection{A semiclassical model}

Next, we proceed to the evolution of the phase.
Some \mwsout{current} descriptions of the temporal coherence of a photon condensate rely on a heuristic phasor model (HPM) \cite{thesisJulian,*thesisJulianEng}, originally developed in laser physics \cite{henry,scully_zubairy_1997}, that we will repeat here with the addition of Kerr interactions. 
According to the HPM, one considers the field to be classical such that the state is defined by a single phasor (corresponding to a coherent state).
The energy of a state with n photons, relative to a situation where all excitations are in the dye, is given by 
 \begin{equation}
 E(n)=\Delta n+\frac{U}{2}n^2,
 \end{equation}where $\Delta$ is the detuning between the cavity and dye transition. As we treat the phasor as an order parameter, the phase oscillates at speed \cite{pitaevski_stringari_2016}
\begin{equation}
v_p=\dv{E}{n}=\Delta+Un.
\end{equation}
As statistical properties remain the same in a rotating frame, we may replace $\Delta$ in numerical simulations by $\Delta'=0$ (no rotation for the vacuum) or $\Delta'=-U\ol{n}$ (no rotation on average) for convergence.
 Absorption, stimulated emission and external losses are here treated as deterministic currents that retain the coherence entirely: they are modeled by
an evolution of the particle number
\begin{equation}\label{eq:hpmdet}
dn=(R-\gamma-\kappa)n\, dt,    
\end{equation}
where $Rn$ is the rate of stimulated emission, $\gamma n$ the rate of absorption and $\kappa n$ the rate of external losses through the mirrors.
Spontaneous emissions into the condensate mode are then taken into account as an additional Poisson process $dM_s$ with expectation $\ol{dM_s}=Rdt$. For each spontaneous emission, a vector $e^{i\phi}$ with unit magnitude and random phase is added to the field $\sqrt{n}e^{i\theta}$, with the motivation that this represents an additional photon that is fully incoherent \cite{stimvsspontem}.  In fact \mwsout{however}, such a spontaneous emission does not deterministically change the photon number with one and may even reduce it.
As before, $R$ and $\gamma$ depend on \mw{$U\ol{n}$ and $M_{\uparrow}$, which is evolved simultaneously with the photon number.} \mw{Since the photon number now varies continuously, the number of molecules in the ground ($M_\downarrow$) and excited ($M_\uparrow$) states are no longer integers.} \mwsout{ As $X$ can only change through external losses and pumping,  $M_{\uparrow} (t)$ and $M_{\downarrow} (t)$ can also no longer be kept as integers.}

\subsection{Predictions}

If $M_\uparrow,M_\downarrow$ are sufficiently large such that $R$ and $\gamma$ remain approximately  constant, the following analytic discussion applies: from geometric reasons, the phase evolves as \cite{henry} 
\begin{equation}
d\theta=\frac{1}{\sqrt{n}}\sin(\phi)\, dM_{\text{spont}}\wv{.}
\end{equation}
From this, we obtain $d\mean(\theta)=0$ and 
\begin{equation}\label{eq:HPMvarphase}
d\Var[\theta]=\ol{\frac{R}{2n}}dt.
\end{equation}
Meanwhile, the photon number evolves as
\begin{equation}
    dn=(R-\gamma-\kappa)n\,dt+(1+2\sqrt{n}\cos\phi)dM_s,
\end{equation}
such that
\begin{equation}\label{eq:HPMmeandens}
d\mean[n]=d\ol{n}=-(\gamma+\kappa) \ol{n}dt+R(\ol{n}+1)dt,
\end{equation}
and
\begin{equation}\label{eq:HPMvardens}
d\Var[n] =2(R-\gamma-\kappa)\Var[n] dt+R(2\ol{n}+1)dt. 
\end{equation}
Despite its simplicity, the HPM is able to give a simple explanation for Schawlow-Townes phase diffusion \cite{scully_zubairy_1997}.

A special case occurs when the particle number vanishes entirely. Here, Eq. \eqref{eq:HPMvarphase} becomes singular. This means that, in absence of other photons, the phase of the spontaneous emission is entirely random over the interval $[0,2\pi[$ and a so-called phase jump occurs \cite{thesisJulian,*thesisJulianEng}. 

\begin{figure*}
   \includegraphics{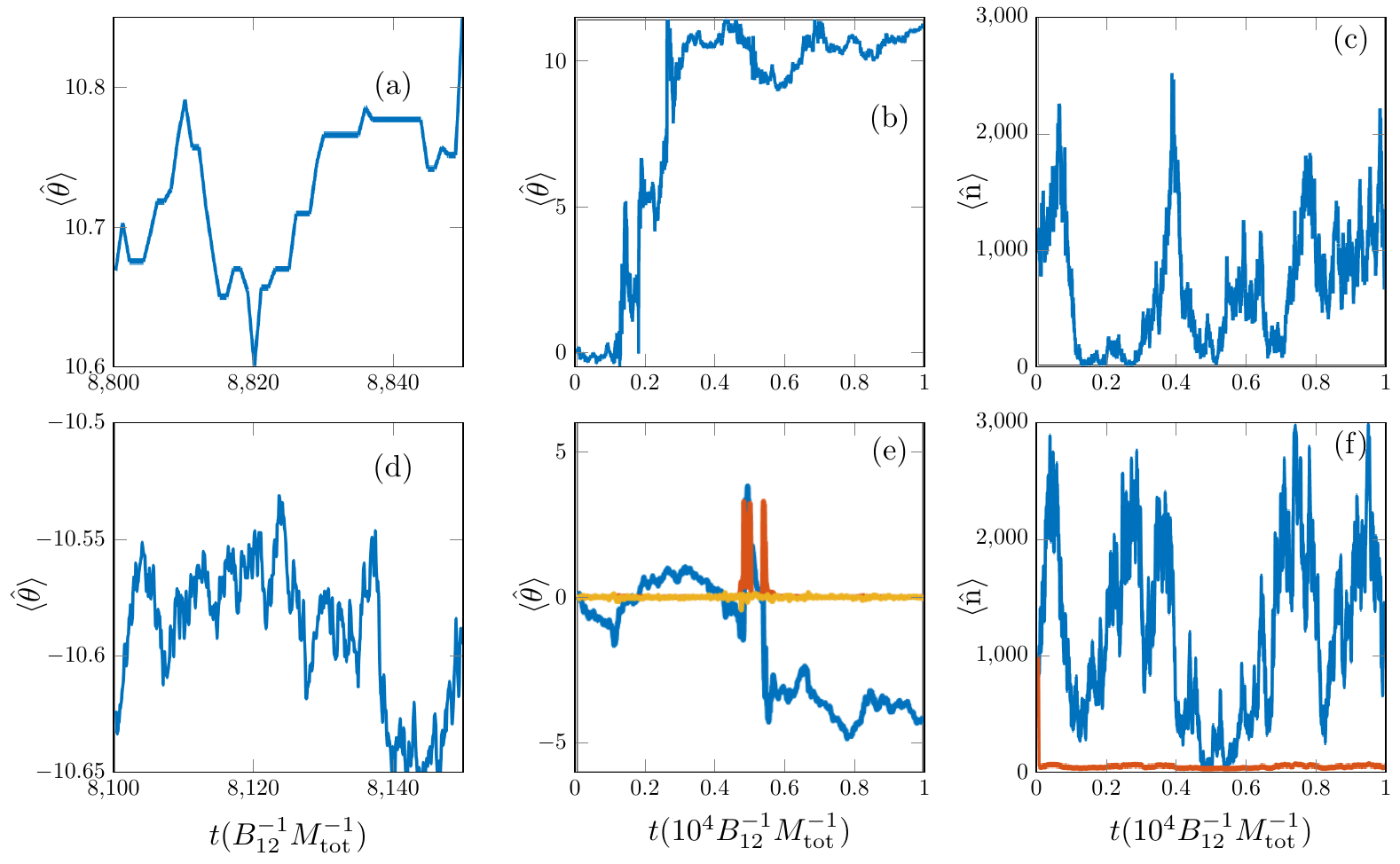}
   \caption{Some typical single shot realizations, as predicted by (a,b,c) the HPM versus (d,e,f) the quantum trajectories. We see that there is a clear qualitative difference on short timescales: whereas the fluctuations of the phasor model (a) are discrete events corresponding to spontaneous emissions, noise remains on all scales within a quantum trajectory (d). However, this qualitative difference is washed away on longer timescales, where the phase evolution according to the HPM (b) becomes equivalent to the one of a trajectory (e, blue line--color online). Also regarding the photon number, the evolution of the HPM (c) is indistinguishable from the trajectory (f,blue line) on \mw{sufficiently large} timescales corresponding to typical experiments. Furthermore, it is clear that both according to the HPM and the trajectories, phase fluctuates the most when the photon number is low. For completeness, we have added also the other Gaussian moments of the trajectory (see Ap. \ref{ap:variational}): on (e) $\edphdph$ (red) and $\edndphsym$ (yellow) are typically small but show spikes at phase jumps. On (f), we see that generally $\edndn$ (red)$<\en$, reflecting number squeezing. Note that $\eph$ is only defined modulo $2\pi$.
   }
    \label{fig:samples}
\end{figure*}

In Figs. \ref{fig:densitydistributions} and \ref{fig:gtwos}, we see that the predictions of the HPM match the exact values for the number statistics, as described in the previous section. In Fig. \ref{fig:samples} (a,b,c) we show the evolution of phase and particle number of a typical HPM sample.

Despite these successes, there are some conceptual difficulties with the HPM.
First of all, by treating the field as classical, any squeezing effects are disregarded. Secondly, the continuous variation of the field is inconsistent with the fact that the emission and absorption are discrete processes at the level of the dye. 
Thirdly, it is physically dubious that only spontaneous emission and neither absorption nor stimulated emission cause shot noise. In the phasor model, the overly large noise from the spontaneous emission actually mimics the shot noise from all loss and gain processes. Note that even though a single spontaneous emission event adds one photon on average, this number has an uncertainty of $\sqrt{n}$, such that a spontaneous emission can even decrease the photon number. 
Finally, regarding the phase, one may wonder what is so special about these spontaneous emission events that these influence $\theta$ whereas other processes do not.

\section{The Quantum Trajectory Description}
\label{sec:TD}

\subsection{A quantum model}
In order to address the questions posed at the end of the previous section, we will study the photon field on a fully quantum mechanical level. Because of its \mw{driven-dissipative} nature, the photon condensate is an open quantum system.
The study of an open system starts with the distinction between system and environment. Here, we will treat the condensate mode as the system to be modelled by a quantum-mechanical stochastic wavefunction $\ket{\psi(t)}$. We will use the notation $\langle\cdot\rangle$ to denote quantum expectation values with respect to this wavefunction. The gain medium on the other hand is modeled classically: of the $M_{\text{tot}}$ dye molecules, the integer amounts $M_\uparrow(t)$($M_\downarrow(t)$) are in the excited(ground-) state as described by stochastic rate equations given in Sec. \ref{sec:numbers}. This is justified because of the frequent collisions with solvent molecules that lead to thermalization \cite{KirtonKeelingThermalizationPhysRevA.91.033826}. 

\begin{figure*}
\includegraphics{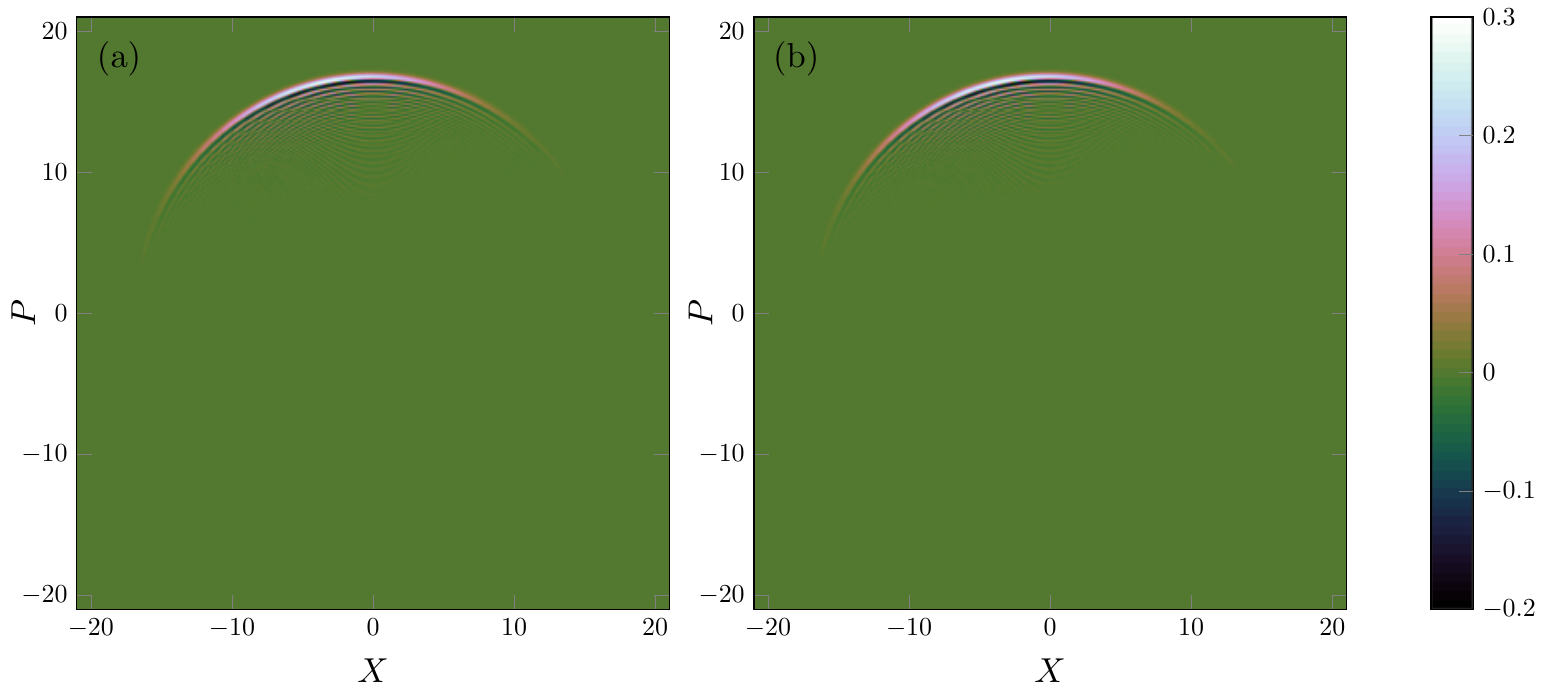}
\caption{Wigner function of a representative trajectory wavefunction in the exact regime, about to make the transition to the variational regime (a). (b): W-function of the corresponding variational trajectory with the same Gaussian moments. The similarity is striking. Colorscheme from \cite{cubehelix}.}
\label{fig:wigner}
\end{figure*}

The coherent evolution of the photons, in the frame rotating at the dye transition frequency, is governed by the Hamiltonian
\begin{equation}
\ham=\Delta \cop\aop+\frac{U}{2}\cop\cop\aop \aop,
\end{equation}
where operators $\aop(\cop)$ annihilate (create) a photon, defining also the number operator $\nop=\cop\aop$. Again, for numerical purposes we can go to a rotating frame and replace $\Delta$ by arbitrary $\Delta'$. In absence of Hamiltonian dynamics, the evolution of the photon field is governed by three processes: gain (corresponding to emission of the dye molecules) occurring at rate $R(\en+1)$, absorption at rate $\gamma\en$ and external losses through the mirrors at rate $\kappa\en$. Because of the discrete nature of the dye excitations, the first two processes are naturally described through a photon-counting unraveling as if the photon number is `measured' by the dye. External losses are modelled by a heterodyne unraveling, firstly because heterodyne detection is typically performed in experiments on this leaking current, and secondly because it keeps the wave function localized in phase space, which will be helpful. By combining these processes, we readily obtain a stochastic Schr\"{o}dinger equation \cite{breuer}
\begin{align}\label{eq:exactev}
\ket{\tilde{\psi}}=&\left[1-i\left(\hat{H}-\frac{i}{2}(\gamma+R+\kappa)\nop\right)dt+\kappa\ea^*\aop dt\right.\nonumber\\&+\left.\sqrt{\kappa}\aop dZ^*+\left(\frac{\aop}{\ea}-1\right)dN+\left(\frac{\cop}{\ea^*}-1\right)dM\right]\ket{\psi}.
\end{align}
Here, the tilde denotes that the left hand side describes the unnormalised wavefunction, to be renormalised after every timestep.   $dZ=\frac{1}{\sqrt{2}}(dW_x+idW_p)$ is a complex (It\^{o}) Wiener noise process such that $|dZ|^2=dt$ and $dN$, $dM$ describe Poisson processes as defined in section \ref{sec:numbers}, where the role of $n$ is now  replaced by $\ev{\nop}$.
Because the photon number typically becomes mesoscopic, exact evolution of \eqref{eq:exactev} in a truncated Fock basis rapidly becomes computationally unfeasible. A number of variational approaches have been proposed for efficient simulation to this extent \cite{daleyreview,wimgutzwiller}. Here, the fact that the system is a single bosonic mode with large occupation, combined with a visual inspection of the Wigner function in Fig. \ref{fig:wigner} leads us to model the field as being Gaussian in particle number and phase: an `$N\Theta$-Gaussian' state as was also used in \cite{gausmethode}. Under this assumption, the field $\ket{\psi}$ is entirely characterized by the expectation values $\en, \eph, \edndn, \edphdph$ and $\edndphsym$, where $\delop_O:=\Oop-\ev{\Oop}$ is defined as the fluctuation of operator $\Oop$. The full variational equations for these expectation values are given in appendix \ref{ap:variational}. 
Note that if external decay vanishes, or if it is described as a photon-counting process,  we find that  $\edndn\rightarrow0,\edphdph\rightarrow\infty$ and the equations reduce reduce to stochastic rate equation as in section \ref{sec:numbers}. Interestingly, it are thus the (small) losses that allow us to define a phase, and avoid the condensate to become a particle number eigenstate \cite{molmernocoherence}. 


Because in the grand-canonical regime the particle number fluctuates also to vanishingly small densities we will use a combination of both the exact \eqref{eq:exactev} and  variational \eqref{eq:densmeaneq}-\eqref{eq:dphcoveq} evolution. That is, we define treshold particle numbers $n_{\text{trans},\searrow}$, $n_{\text{trans},\nearrow}$. When $\en<n_{\text{trans},\searrow}=200$ we perform the exact evolution and when $\en>n_{\text{trans},\nearrow}=240$ we perform the variational evolution. In the intermediate window $n_{\text{trans},\searrow}<\en<n_{\text{trans},\nearrow}$, the method of the previous regime remains in use. The transitions between methods are discussed in more detail in Appendix \ref{ap:transitions}. Whereas the variational equations are solved by a straightforward Euler method with direct addition of the Poisson increments, for proper numerical convergence of the exact trajectories, an approach where the deterministic and diffusive evolution is separated from the individual jumps \cite{jumpdiffusion} is used.

\subsection{Predictions}
Similar to the HPM, we calculate the evolution of the moments of the field while treating $\gamma,R$ as constant. We will use the variational equations from Ap. \ref{ap:variational} as a starting point, considering highest order $\mathcal{O}(\en)=\mathcal{O}(\edndn)=\mathcal{O}(\edphdph^{-1})$ (this relation between orders can be seen from a coherent state where $\en=\edndn$ and  relation \eqref{eq:ptydensphase}). Under the assumption $\edndphsym\equiv0$ and neglecting noise, one obtains as stationary solution for the phase variance \eqref{eq:phasevareq},
\begin{equation}\label{eq:statphasevar}
\edphdph_{\text{stat}}=\frac{1+\sqrt{2(1+\frac{\gamma+R}{\kappa})}}{4\en}.
\end{equation}
Substituting in \eqref{eq:phasemeaneq} results in
\begin{equation}
d\eph=\frac{\kappa+\gamma+R}{2\sqrt{\en}}dW_\theta,
\end{equation}
where $dW_\theta$ denotes a real Wiener process from which
\begin{equation}\label{eq:TDmeanphase}
d\mean[\phop]=d\ol{\eph}=0
\end{equation} 
and
\begin{equation}\label{eq:TDvarphase}
d\intervar{\phop}=d\left[\ol{\phop\phop}\right]-d\left[\ol{\phop}^2\right]=\ol{\frac{\kappa+\gamma+R}{4\en}}dt.
\end{equation}
In trajectory simulations, the total variance of an observable $\Var[\Oop]=\intravar{\Oop}+\intervar{\Oop}$ can be decomposed into the intra-trajectory variance $\intravar{\Oop}=\ol{\ev{\delta_{\Oop}\delta_{\Oop}}}$ and the inter-trajectory variance $\intervar{\Oop}=\Var\ev{\Oop}$ \cite{breuer}. Here, from our stationary assumption, $d\intravar{\phop}=0$ such that $d\Var[\phop]=d\intervar{\phop}$. 

Similarly, we obtain the stationary solution for the particle number from \eqref{eq:densvareq}:
\begin{equation}
\edndn_{\text{stat}}=\frac{\sqrt{\kappa}}{\sqrt{2}\sqrt{\gamma+R}}\en,
\end{equation}
where we have used $\kappa\ll\gamma,R$.
Substituting in \eqref{eq:densmeaneq} results in
\begin{equation}\label{eq:TDmeandens}
d\mean[\en]=d\ol{\en}=(R-\gamma-\kappa)\ol{\en} dt
\end{equation}
and
\begin{align}\label{eq:TDvardens}
d\Var{\nop}&=d\intervar{\nop}=d\ol{\en^2}-d(\ol{\en}^2)\nonumber\\&=[2(R-\gamma-\kappa)\Var{\nop}+(\gamma+R+\kappa)\ol{\en}]dt.
\end{align}

Comparing Eqs. \eqref{eq:TDmeanphase},\eqref{eq:TDvarphase},\eqref{eq:TDmeandens},\eqref{eq:TDvardens} with the predictions from the HPM \eqref{eq:HPMvarphase},\eqref{eq:HPMmeandens},\eqref{eq:HPMvardens} the similarity is evident. In line with the expectations, the HPM predicts a variance increase proportional to the gain $R$, whereas this is symmetrical in $R,\gamma,\kappa$ according to our trajectory result. In the steady-state however, $R\approx\gamma+\kappa$ at least to highest order in $\en$, such that the time-averaged result is the same.
This correspondence is also reminiscent to the result for a laser \cite{scully_zubairy_1997}: also there, the phasor model predicts a phase diffusion proportional to the gain coefficient $R$ whereas a more detailed quantum-mechanical derivation makes clear that $\frac{R+\kappa}{2}$ is the correct quantity, although both become equivalent at threshold where $R\approx\kappa$.

In Fig. \ref{fig:samples} (d,e,f) a representative evolution of phase and particle number during a single trajectory simulation is shown. On short times there is a qualitative difference with the HPM: discrete steps corresponding to emission events are replaced by scale-invariant noise, induced by the heterodyne detection and fueled by all emission and absorption processes. However, current experiments cannot resolve these small fluctuations on short timescales, as long as they do not affect correlation functions. It is clear from fig. \ref{fig:samples} that the predictions of both methods become qualitatively indistinguishable on longer timescales. We see also that for both methods, phase diffuses more rapidly at lower photon numbers, as predicted by Eqs. \eqref{eq:HPMvarphase},\eqref{eq:TDvarphase}. 

\section{First order coherence}
\label{sec:1coherence}
In Fig. \ref{fig:gones} the first order correlation function $g^{(1)}(\tau)=\frac{\overline{\alpha^*(t)\alpha(t+\tau)}}{\overline{\abs{\alpha(t)}^2}}$ is shown, and again predictions of the HPM agree well with the exact result, obtained by trajectories. In the noninteracting case, the decay of correlations is clearly exponential, whereas it is Gaussian in presence of interactions.  The decay can be understood from multiple points of view, let us start with the noninteracting case.

\subsection{Schawlow-Townes effect from the canonical to the grandcanonical regime: the influence of `phase jumps'}

According to the HPM picture, absorption and stimulated emission only affect the phasor radially, as $d\sqrt{n}=\frac{dn}{2\sqrt{n}}=\frac{R-\gamma}{2}\sqrt{n}\, dt$, where we used eq. \eqref{eq:hpmdet}, and neglected the external losses.
The stochastic evolution of the HPM can then be written in terms of the phasor alone as
\begin{equation}
d\alpha=-\frac{\gamma-R}{2}\alpha\,dt+dS,
\end{equation}
where $dS$ is additive noise corresponding to the spontaneous emissions. In the grandcanonical limit, $\gamma$ and $R$ can be treated as constants, such that for the expectation value,
\begin{equation}
d\ol{\alpha}=-\frac{\gamma-R}{2}\ol{\alpha}=\frac{-R}{2\ol{n}}\ol{\alpha}.
\end{equation}
From the quantum regression theorem \cite{scully_zubairy_1997}, it is then clear that
\begin{equation}
g^{(1)}(\tau)=e^{-t/\tau_c^{(1)}},
\end{equation}
where 
\begin{equation}\label{eq:gcg1decayrate}
\frac{1}{\tau_c^{(1)}}=\frac{R}{2\ol{n}}=\frac{B_{21}M_{\uparrow}}{2\ol{n}}\approx\frac{B_{12}M_{\downarrow}}{2\ol{n}}=\frac{B_{12}M_{\text{tot}}}{2\ol{n}(1+e^{\beta\Delta})}.
\end{equation}

\begin{figure}
\includegraphics[width=\columnwidth]{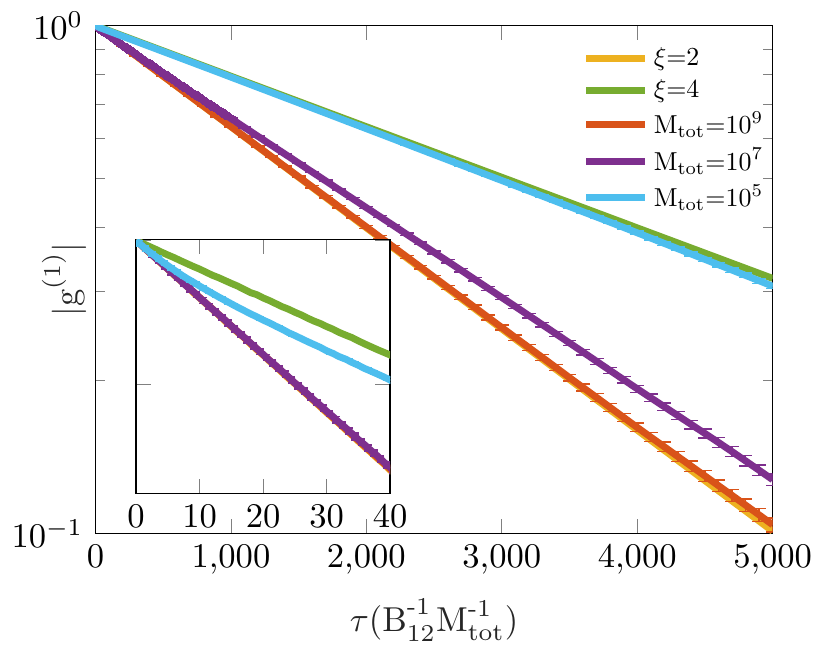}
     \caption{$g^{(1)}(\tau)$ in absence of interactions, for different values of $M_\text{tot}$ ($10^9$-red, $10^7$-purple,$10^5$-cyan) and hence $M_\text{eff}$, together with the asymptotic exponential decays $\xi=2$ (yellow) and $\xi=4$ (green). Again, the other parameters are as in Tab. \ref{tab:parameters} and $\ol{n}=1000$. Inset: even towards the canonical regime, there is a short initial timespan of grananonical ($\xi=4$) decay of the order of  $\tau_c^{(2)}$. 
   }
    \label{fig:g1asfunctionofreservoir}
\end{figure}

By another line of reasoning, the phase evolution is dominated by large `phase jumps' when the photon number in the cavity vanishes as described in \cite{Schmittphasecoherence,thesisJulian,*thesisJulianEng}. From estimating the probability of having zero photons in the cavity, one obtains the `phase jump rate'
$\Gamma_{\text{PJ}}^0=\frac{B_{12}M_\downarrow}{\ol{n}^\zeta}$. In the grandcanonical limit, $\zeta=1$, so that this is consistent with \eqref{eq:gcg1decayrate} up to a scaling factor of order one.  The picture of phase jumps would further predict that the phase evolution is suppressed if the probability for the zero-photon state vanishes ($\zeta\rightarrow\infty$), as occurs outside of the grandcanonical limit, towards the canonical regime $M_{\text{eff}}\ll\ol{n}^2$, where $M_\text{eff}$ is defined as in Eq. \eqref{eq:meff}. By additional numerical simulations shown in Fig. \ref{fig:g1asfunctionofreservoir}, we see that this is not the case. At most, the first-order coherence time, rescaled with the molecule number, only scales by a factor two. This is not entirely unsurprising as the remaining value corresponds to the standard Schawlow-Townes dephasing \cite{scully_zubairy_1997} occurring in the limit where the number distribution is sharply peaked, as can already be seen from \eqref{eq:HPMvarphase}. 
We are thus led to write generically
\begin{equation}\label{eq:g1general}
\Gamma_1 \equiv \frac{1}{\tau_c^{(1)}}=\frac{B_{12}\ol{M_{\downarrow}}}{\xi\ol{n}},
\end{equation}
where $\xi=2$ in the grandcanonical limit and $\xi=4$ in the canonical limit. Comparing with the decay of second-order coherence \eqref{eq:interactingtc2}, we find that
\begin{align}\label{eq:timeratio}
\frac{\tau_c^{(1)}}{\tau_c^{(2)}}&=\frac{\Gamma_2}{\Gamma_1}\\
&=\xi\left[1+\sigma+\frac{\ol{n}^2}{M_{\text{eff}}}\right]\\
&=\frac{\xi}{\eta^2}=\frac{\xi}{g^{(2)}(0)-1}, 
\end{align}
where we have used \eqref{eq:alggtwo0} for the third equality.
However, on the inset of fig. \ref{fig:g1asfunctionofreservoir}, we see that on sufficiently short timescales, there is always an initial decay with $\xi=2$. There is a clear intuition here. In general, $\gamma-R=\ol{\gamma}-\ol{R}+(B_{12}+B_{21})\delta n$, where $\delta n=n-\ol{n}$. When, at short timescales, $\delta n$ remains approximately constant, a similar reasoning to the grandcanonical regime yields
\begin{align}
g^{(1)}(\tau)&=\frac{e^{-\ol{R}\tau/(2\ol{n})}}{\ol{n}}\ol{n\exp\left(-\frac{(B_{12}+B_{21})\delta n}{2}\tau\right)}\nonumber\\&\approx e^{\frac{-\ol{R}}{(2\ol{n})}\tau}\left(1-\frac{\eta^2}{2}(B_{12}+B_{21})\ol{n}\tau\right)\nonumber\\&\approx e^{\frac{-\ol{R}}{(2\ol{n})}\tau}.
\end{align}
At later times, higher-order effects set in and restrict the dynamics.
It is instructive to compare these results with the ones obtained by Whittaker and Eastham \cite{Whittakerpolaritons1,Whittakerpolaritons2} in a polariton context. There, a Schawlow-Townes decay is predicted to be of the form

\begin{equation}\label{eq:WhittakerST}
\abs{g^{(1)}(\tau)}=\exp\left[\frac{\eta^2}{4}(e^{-\Gamma_2\tau}-\Gamma_2\tau-1)\right].
\end{equation}
For $\Gamma_2\tau\ll 1$, this reduces to an exponential decay at rate $\frac{\eta^2}{2}\Gamma_2$, whereas  for $\Gamma_2\tau\gg 1$, \eqref{eq:WhittakerST} decays exponentially at rate $\frac{\eta^2}{4}\Gamma_2$. This can be understood because at short times the number fluctuations contribute to the decay of $g^{(1)}$, but after a time $1/\Gamma_2$ only Schawlow-Townes phase diffusion remains. 
There remain a few differences between the physics of \eqref{eq:WhittakerST} and the photon condensate. First, \eqref{eq:WhittakerST} is derived in \cite{Whittakerpolaritons1,Whittakerpolaritons2} under the explicit assumption that the whole ensemble has a Gaussian number distribution \mw{peaked} around $\ol{n}$  \footnote{This is different from our ansatz where we consider only individual trajectories as having a Gaussian number distribution.}. This assumption is physical for the photon condensate only in the canonical regime, where the probability of having zero photons is negligible. \mw{This means that the predicted long time value of $\xi$ equals 4, in agreement with our prediction for the canonical regime.}

\mwsout{In an intermediate regime, we consequently see that $\xi$ increases by less than a factor two as we can see on Fig. \ref{fig:g1asfunctionofreservoir}.}

\mw{Finally}, the timescale of the transition between the two decay rates \mw{is for photon condensates}  not determined by $\Gamma_2$\mwsout{itself}, but only by the second term in \eqref{eq:interactingtc2}, which is responsible for the time-dependence of $R$ and $\gamma$. In the grandcanonical limit, this means that a slowing of the decay would only take place at infinitely long times unlike the prediction of \eqref{eq:WhittakerST}, as \mw{is seen} on Fig. \ref{fig:gones}.

\subsection{Effect of Kerr-interactions}

\begin{figure}
\includegraphics[width=\columnwidth]{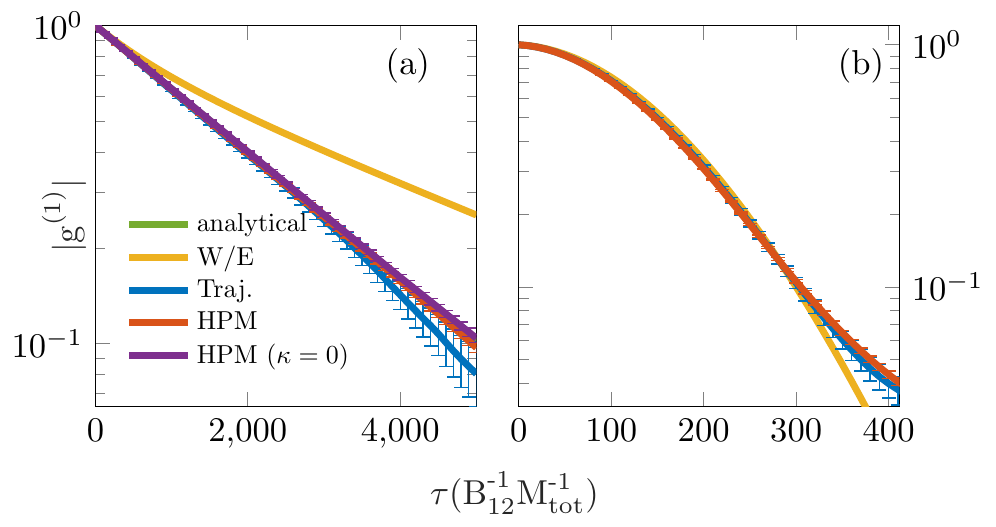}
    \caption{First order correlation function for the parameters of Tab \ref{tab:parameters}, (a) noninteracting and (b) interacting case. Colors as before, Yellow: the polariton result of \eqref{eq:WhittakerST} and \eqref{eq:WhittakerHenry} from \cite{Whittakerpolaritons1,*Whittakerpolaritons2}.  The result without losses(purple) is also obtained by the HPM. The quantum-trajectory results are obtained from $84$(noninteracting) or $112$ (interacting) independent samples, each evolving a time $10^4 B_{12}^{-1}M_\text{tot}^{-1}$}
    \label{fig:gones}
\end{figure}

In presence of finite photon-photon interactions, the decay of first-order coherence is altered, as we see on the right panel of Fig. \ref{fig:gones}. The profile is rather Gaussian than exponential. This is characteristic for the so-called Henry mechanism \cite{henry}. Whereas Schawlow-Townes decay is attributed to direct fluctuations of the phase, Henry decay results from phase diffusion as a consequence of number fluctuations causing a change of the interaction energy. For the Henry effect, Ref. \cite{Whittakerpolaritons1,*Whittakerpolaritons2} predicts an additional decay
\begin{equation}\label{eq:WhittakerHenry}
\abs{g^{(1)}(\tau)}=\exp\left[-\frac{\eta^2U^2}{\Gamma_2^2}(e^{-\Gamma_2\tau}+\Gamma_2\tau-1)\right].
\end{equation}
which reduces to a Gaussian decay with characteristic time $(\sqrt{2}/U\ol{\en}\eta)$ at short timescales. As we see on figure \ref{fig:gones}, there is good agreement with our numerical results on short times, although deviations occur at later times that we can attribute to the non-Gaussian character of the number distribution.

The fact that the prediction of Gaussian decay for short timescales remains valid in the grandcanonical limit can be understood because in a frame rotating at \mw{the} bare cavity frequency, the expectation value of a phasefactor of a state with $n$ photons at time $t$ is

\begin{align}
\ol{e^{(-iUnt)}}&=\exp\left[\sum_{m=1}^\infty k_m\frac{(-i)^mU^m t^m}{m!}\right]\nonumber\\&=e^{-iU\ol{n}t}e^{\frac{-1}{2}U^2t^2(\ol{n^2}-\ol{n}^2)+\mathcal{O}(Ut)^3},
\end{align}
where $k_m$ is the $m$th cumulant of the distribution of $n$. Here, we implicitly assumed that $n$ remains approximately constant on short times.

\section{Conclusions}
\label{sec:conclusions}
We have studied the temporal coherence of a single\wv{-}mode photon condensate both with and without weak Kerr interactions, and have also accounted for the effects of driving and dissipation. We have shown thermodynamically and dynamically how interactions reduce the number fluctuations and calculated the corresponding enhancement of the decay of second-order correlations, which we numerically verified by stochastic rate equations. The driven-dissipative nature of a realistic photon condensate causes $g^{(2)}(\tau)<1$ at large $\tau$, because of the coupling of the photon number to the slow reservoir dynamics. We have reviewed the heuristic phasor model, and shown analytically and numerically its adequacy in describing experimentally relevant quantities while being numerically efficient. For comparison, a quantum mechanical model of the condensate mode based on variational quantum trajectories was introduced, assuming only classical coupling with the dye-molecules. Exhibiting similarities with both a laser far from equilibrium \cite{henry} and an isolated atomic BEC~\cite{BECcoherence},
we have observed both Schawlow-Townes (exponential) and Henry (Gaussian) contributions to the decay of phase correlations in the photon condensate, depending on the interaction strength. As the effect of `phase jumps' is similar to the standard Shawlow-Townes effect, we found no qualitative differences in the coherence between canonical and grandcanonical regimes. We have shown how first- and second-order coherence times are related by the number fluctuations. An interesting open question is to what extent the picture above changes in presence of thermo-optical effects. To study the latter, longer evolution times are necessary, but as we have shown an approach based on the numerically efficient heuristic phasor model is likely to be sufficient. 
Another possible extension is the study of coupled condensate modes, as appear in lattices of cavities. Also within a single cavity, there can be interesting physics arising from mode competition for some parameter values \cite{nonmarkovsimulation}, which we have omitted here.

Finally, we have assumed here that the absorption and emission processes are Markovian and coupling between the photons and molecules is weak. 

It remains an open question to what extent our conclusions remain valid outside of these approximations \cite{RevModPhys.nonmarkov}. Because our conclusions in strong thermal equilibrium and weak losses are \mw{similar} \mwsout{not unsimilar} to a threshold laser far from equilibrium, both regarding the validity of the HPM and the shape of the autocorrelation functions, we expect these conclusions to remain valid for all usual parameter values for photon condensates.

\begin{acknowledgments}
We acknowledge stimulating discussions with J. Schmitt, M. Weitz, F. \"Ozt\"urk and H. Kurkjian. We furthermore thank J. Schmitt and M. Weitz for their feedback on our manuscript. This work was financially supported by the FWO Odysseus program. Part of the computational resources and services used in this work were provided by the VSC (Flemish Supercomputer Center), funded by the Research Foundation - Flanders (FWO) and the Flemish Government – department EWI.
\end{acknowledgments}

\appendix

\begin{widetext}
\section{The variational evolution}\label{ap:variational}
Following reference \cite{gausmethode} (which has however a different sign convention for the parameter $\Delta$), the trajectory evolution \eqref{eq:exactev} can be recast in terms of (normalized) expectation values $\ev{\Oop}$ to be

\begin{align}\label{eq:correv}
d\ev{O}=&i\langle\comm{H}{O}\rangle dt-\frac{\gamma+R+\kappa}{2}\left(\langle\nop O\rangle+\langle O\nop\rangle\right)dt+(\gamma+R)\ev{O}\en dt+\kappa\langle\cop O\aop\rangle dt\nonumber\\&+\sqrt{\kappa}\left(\langle\cop\hat{\delta_O}\rangle dZ+\langle\hat{\delta_O}\aop\rangle dZ^*\right)+\left(\frac{\langle\cop O \aop\rangle}{\en}-\langle O\rangle\right)dN+\left(\frac{\langle\aop O \cop\rangle}{\en+1}-\langle O\rangle\right)dM,
\end{align}
where again $\hat{\delta_O}=\hat{O}-\ev{O}$.
By introducing a Dirac phase through $\aop=:e^{i\hat{\theta}}\sqrt{\nop}$ satisfying $[\nop,\phop]=i$ and using Wick's theorem, we obtain for the Gaussian correlation functions $\en, \eph, \edndn=\enn-\en^2, \edphdph=\ephph-\eph^2,\edndphsym=\enph/2+\ev*{\phop\nop}/2-\en\eph$ the evolution  

\begin{align}
d\en=&-(\gamma+R)\edndn dt-\kappa\en dt+2\Re\left[(C_2-C_1)\sqrt{\kappa}dZ\right]\nonumber\\&+\left(\frac{\edndn}{\en}-1\right)dN+\left(\frac{\edndn}{\en+1}+1\right)dM\label{eq:densmeaneq}\\
d\edndn=&-2\kappa\edndn dt+\kappa\en dt-2\kappa\abs{C_2-C_1}^2 dt\nonumber\\&+2\Re\left[(D_3-2C_2+C_1(1-\edndn))\sqrt{\kappa}dZ\right]\nonumber\\&-\frac{\edndn^2}{\en^2}dN-\frac{\edndn^2}{(\en+1)^2}dM\label{eq:densvareq}\\
d\eph=&\left(-\Delta+\frac{U}{2}\right)-U\en-(\gamma+R)\edndphsym dt+2\Re\left[C_6\sqrt{\kappa}dZ\right]\nonumber\\&+\frac{\edndphsym}{\en}dN+\frac{\edndphsym}{\en+1}dM\label{eq:phasemeaneq}\\
d\edphdph=&-2U\edndphsym+\frac{\kappa}{4}E_1 dt-2\kappa\abs{C_6}^2 dt+2\Re\left[(D_1-\edphdph C_1)\sqrt{\kappa} dZ\right]\nonumber\\&+\left(\frac{-\edndphsym^2}{\en^2}+\frac{E_1}{4\en}\right)dN+\left(\frac{-\edndphsym^2}{(\en+1)^2}+\frac{E_2}{4(\en+1)}\right)dM\label{eq:phasevareq}\\
d\edndphsym=&-U\edndn-\kappa\edndphsym dt-2\kappa\Re\left[(C_2-C_1)C_6^*\right]dt\nonumber\\&+2\Re\left[\left(-C_6-\left(\edndphsym+\frac{i}{2}\right)C_1+D_2\right)\sqrt{\kappa}dZ\right]\nonumber\\&-\edndn\frac{\edndphsym}{\en^2}dN-\edndn\frac{\edndphsym}{(\en+1)^2}dM\label{eq:dphcoveq}.
\end{align}

Here, the coefficients $C$ and $D$ are defined as in \cite{gausmethode} and

\begin{align}
E_1:=&\ev{\frac{1}{\nop}}\approx\frac{1}{\en}\left(1+\frac{\edndn}{\en^2}\right)\\
E_2:=&\ev{\frac{1}{\nop+1}}\approx\frac{1}{\en+1}\left(1+\frac{\edndn}{(\en+1)^2}\right).
\end{align}

By considering the measurements to be perfect, we can do a restriction towards pure states, for which relation
\begin{equation}\label{eq:ptydensphase}
\edndn\edphdph-\edndphsym^2=\frac{1}{4}
\end{equation}
is satisfied \cite{gausmethode}. This constraint allows to compute $\edphdph$ (or another variance) directly from the others.

\section{Transitions between exact and variational regimes} \label{ap:transitions}
\subsection{From variational to exact}

When $\en$ decreases from the variational regime below the threshold $n_{\text{trans},\searrow}$, the $N\Theta$-Gaussian state must be explicitly expressed in Fock-base to continue numerically exact evolution.

A generic Gaussian density operator \cite{quantumnoise} can, by definition, be written as

\begin{equation}
\hat{\rho}=\mathcal{N}\exp{-\beta\Hat{H}_{\text{eff}}}    
\end{equation}
for a quadratic $\Hat{H}_{\text{eff}}$. A Gaussian state that is pure is obtained by taking the limit $\beta\rightarrow\infty$, which is equivalent with taking the lowest eigenvector (`ground state') of $\Hat{H}_{\text{eff}}$.

In order to construct $\Hat{H}_{\text{eff}}$, we need explicit matrix representations of $\nop$ and $\phop$. For the particle number operator $\nop=\cop\aop$ this is straightforward. Regarding $\phop$ we encounter the fact that phase is no true observable, with the consequence that, strictly speaking, no hermitian phase operator exists \cite{phaseopreview}. However, as long as the Fock space is truncated (which is the case here) at level $N_{\text{max}}$ , a meaningful phase operator can be obtained through the Pegg-Barnett formalism to be

\begin{equation}
    \phop_{PB}(\theta_0)=\theta_0+\frac{N_{\text{max}}\pi}{N_{\text{max}}+1}+\frac{2\pi}{N_{\text{max}}+1}\sum_{j\neq k}^{N_{\text{max}}}\frac{\exp{i(j-k)\theta_0}\ket{j}\bra{k}}{\exp{i(j-k)2\pi/(N_{\text{max}}+1)}-1} 
\end{equation}
\cite{phaseopreview}. One free parameter, $\theta_0$, remains, corresponding to the phase-cut. That is, because phase is a periodic variable, there must be a cut where (going counterclockwise) the phase sharply changes with $-2\pi$. In order to avoid secondary effects of this cut, we will use $\theta_0=\eph-\pi$ such that the $N\Theta$-Gaussian state is as far from the phase-cut as possible.

Using the operators $\nop$ and $\phop:=\phop_{PB}(\eph-\pi)$, an effective Hamiltonian
\begin{equation}
\Hat{H}_{\text{eff}}=(\nop-\en)^2\edphdph+(\phop-\eph)^2\edndn-(\nop-\en)(\phop-\eph)\edndphsym-(\phop-\eph)(\nop-\en)\edndphsym
\end{equation}
is constructed.
$\ket{\psi}$ is obtained as the eigenvector of $\Hat{H}_{\text{eff}}$ corresponding to the lowest eigenvalue.

\subsection{From exact to variational}
When, in the exact regime, $\en=\bra{\psi}\nop\ket{\psi}$ increases above $n_{\text{trans},\nearrow}$, the Gaussian moments must be computed. As in the previous case, this is entirely straightforward regarding $\en$ and $\edndn$. For the phase, we want to use an operator $\phop_{PB}(\theta_0)$ again, although it is this time not a priori clear which value of $\theta_0$ to use. We will therefore do an initial guess $\theta_0^{(0)}$ and construct a corresponding $\phop_{PB}^{(0)}=\phop_{PB}(\theta_0^{(0)})$. We then iteratively use $\theta_0^{i+1}=\bra{\psi}\phop_{PB}^{(i)}\ket{\psi}-\pi$ and repeat the procedure self-consistently until convergence is reached. The resulting $\phop_{PB}(\theta_0^{(f)})$ can then be used to proceed in calculating the Gaussian moments. We have verified that the above procedure results in the expected phase $\arg{\left[\alpha\right]}$ when applied to an arbitrary coherent state $\ket{\alpha}$.

After this transition, we use purity relation \eqref{eq:ptydensphase} as a numerical check and as verification for the validity of the Gaussian ansatz.

The fact that in this work two different definitions of a phase operator (Dirac and Pegg-Barnett) are used causes no problems because the particle number at the transitions is sufficiently large and the trajectory states are sufficiently well-behaved.

\end{widetext}


%

\end{document}